\documentclass[aps,prd,preprint,amsmath,amssymb,showpacs,preprintnumbers,superscriptaddress]{revtex4-2}
\usepackage{graphicx}
\usepackage{dcolumn}
\usepackage{bm}
\usepackage{color}
\usepackage{hyperref}
\usepackage{ulem}
\usepackage{mathrsfs}
\usepackage{braket}

\allowdisplaybreaks[4]

\begin{document}

\title{Exotic spin-dependent interactions through unparticle exchange}

\author{L. Y. Wu}
\affiliation{Institute of Nuclear Physics and Chemistry, CAEP, Mianyang, Sichuan 621900, China}
\affiliation{Key Laboratory of Neutron Physics, Institute of Nuclear Physics and Chemistry, CAEP, Mianyang 621900, Sichuan, China}

\author{K. Y. Zhang}\email{zhangky@caep.cn}
\affiliation{Institute of Nuclear Physics and Chemistry, CAEP, Mianyang, Sichuan 621900, China}
\affiliation{Key Laboratory of Neutron Physics, Institute of Nuclear Physics and Chemistry, CAEP, Mianyang 621900, Sichuan, China}

\author{H. Yan}\email{h.y\_yan@qq.com}
\affiliation{Institute of Nuclear Physics and Chemistry, CAEP, Mianyang, Sichuan 621900, China}
\affiliation{Key Laboratory of Neutron Physics, Institute of Nuclear Physics and Chemistry, CAEP, Mianyang 621900, Sichuan, China}

\begin{abstract}
  The potential discovery of unparticles could have far-reaching implications for particle physics and cosmology. For over a decade, high-energy physicists have extensively studied the effects of unparticles. In this study, we derive six types of nonrelativistic potentials between fermions induced by unparticle exchange in coordinate space. We consider all possible combinations of scalar, pseudo-scalar, vector, and axial-vector couplings to explore the full range of possibilities. Previous studies have only examined scalar-scalar (SS), pseudoscalar-pseudoscalar (PP), vector-vector (VV), and axial-axial-vector (AA) type interactions, which are all parity even. We propose SP and VA interactions to extend our understanding of unparticle physics, noting that parity conservation is not always guaranteed in modern physics. We explore the possibilities of detecting unparticles through the long-range interactions they may mediate with ordinary matter. Dedicated experiments using precision measurement methods can be employed to search for such interactions. We discuss the properties of these potentials and estimate constraints on several coupling constants based on existing experimental data. Our findings indicate that the coupling between vector unparticles and fermions is constrained by up to 9 orders of magnitude more tightly than the previous limits.
\end{abstract}

\date{\today}

\maketitle

\section{Introduction}

The concept of symmetry has a profound impact on the development of modern physics.
In 2007, Georgi~\cite{Georgi2007PRL} proposed the existence of a hidden conformal symmetry sector beyond the Standard Model (SM) of particle physics, termed unparticle.
Since then, the effects of unparticles have been explored in many subjects, including collider physics~\cite{Georgi2007PLB, Cheung2007PRL, CMS2015EPJC, Aliev2017PRD, Soa2018NPB}, neutrino physics~\cite{Montanino2008PRD, Pandey2019JHEP}, cosmology and astrophysics~\cite{Davoudiasl2007PRL, Das2008PRD}, quantum electrodynamics~\cite{Liao2007PRD, Frassino2017PLB}, dark energy~\cite{Dai2009PRD, Artymowski2021PRD(L), Abchouyeh2021PRD, Artymowski2022PRD}, etc.
The recent nonrelativistic extension of unparticle physics has led to a new concept of unnucleus~\cite{Hammer2021PANS} and provided an interpretation of neutral charm mesons near threshold~\cite{Braaten2022PRL}.

One of the most intriguing phenomenologies of unparticle physics is the possibility of exotic long-range interactions mediated by unparticles~\cite{Liao2007PRL}.
To date, only electromagnetic and gravitational interactions have been observed as long-range interactions.
However, considerable efforts~\cite{Moody1984PRD, Dobrescu2006JHEP, Costantino2020JHEP, Ding2020PRL, Wu2022PRL, Liang2022NSR, Wei2023PRL} have been devoted to searching for extra long-range interactions, which could be mediated by axions~\cite{Wilczek1978PRL, Weinberg1978PRL}, paraphotons~\cite{Holdom1986PLB, Dobrescu2005PRL}, $Z'$ bosons~\cite{PDG2020PTEP}, and graviphotons~\cite{Atwood2000PRD}.
The discovery of new long-range interactions would have a tremendous impact on our understanding of nature, such as the strong $CP$ problem \cite{Kim2010RMP} and dark matter \cite{Braaten2019RMP}.

Liao and Liu~\cite{Liao2007PRL} first proposed the idea of unparticle-mediated long-range interactions, using experimental data on the long-range spin-dependent interaction of electrons to constrain the couplings of unparticles to electrons.
It was also noted that unparticles could mediate spin-independent long-range interactions, which might modify the inverse square law of gravity~\cite{Goldberg2008PRL, Deshpande2008PLB, Bertolami2009PRD}, contribute to the ground-state energy of the hydrogen atom~\cite{Wondrak2016PLB}, and influence the measurement of spacetime curvature near Earth~\cite{Poddar2022EPJC}.
The possibility of a long-range interaction between neutrinos from unparticle exchange was also explored in solar neutrino experiments~\cite{Gonzalez-Garcia2008JCAP}.
Only long-range interactions originating from the same type of vertex, such as scalar-scalar (SS), pseudoscalar-pseudoscalar (PP), vector-vector (VV), and axial-axial-vector (AA), were considered in these works.
Different types of couplings between unparticles and SM particles, such as scalar-pseudoscalar (SP) and vector-axial-vector (VA) interactions, could also exist.
The interactions of SS, PP, VV, and AA types conserve all of the discrete symmetries, including charge conjugation ($C$), parity ($P$), and time-reversal ($T$).
In contrast, the SP-type interaction violates both $P$ and $T$, while the VA-type interaction violates $P$ and $C$.
It is important to note that the conservation of discrete symmetries such as $P$ and $CP$ cannot be assumed in modern physics.
Taking such interactions into account when exploring extra long-range interactions or constraining them with existing experimental data could deepen our understanding of unparticle physics and provide new insights into new physics searches.

In this paper, we derive the interactions between fermions from unparticle exchange, considering all possible combinations of scalar, vector, pseudoscalar, and pseudovector couplings.
We present six types of potentials explicitly in coordinate space under the nonrelativistic limit.
We discuss the properties of the derived potentials and estimate the constraint on some couplings based on existing experimental data.

\section{Long range spin-dependent interactions via unparticle exchange}

Following the scenario described in Ref.~\cite{Georgi2007PRL}, at some high energy scale, the SM fields can interact with the $\mathcal{BZ}$ (Banks-Zaks~\cite{Banks1982NPB}) field having a nontrivial infrared fixed point through the exchange of messenger particles with a large mass scale $M_\mathcal{U}$.
Below this mass scale, nonrenormalizable couplings are suppressed by inverse powers of $M_\mathcal{U}$ and take the generic form of
$\frac{1}{M_\mathcal{U}^k}\mathcal{O}_{\mathrm{SM}}\mathcal{O}_{\mathcal{BZ}}$,
where $\mathcal{O}_{\mathrm{SM}}$ and $\mathcal{O}_{\mathcal{BZ}}$ denote operators built out of SM and $\mathcal{BZ}$ fields, respectively.
The renormalizable couplings of the $\mathcal{BZ}$ fields then induce dimensional transmutation, and the scale-invariant unparticle fields emerge at another energy scale $\Lambda_\mathcal{U}$.
Below the scale $\Lambda_\mathcal{U}$, $\mathcal{BZ}$ operators match onto unparticle operators $\mathcal{O}_\mathcal{U}$, and the unparticle interaction with SM particles at low energy has the form
\begin{equation}
\mathcal{L}_{\mathrm{int}}\propto \frac{1}{\Lambda_\mathcal{U}^{d_\mathcal{U}+d_\mathrm{SM}-4}}\mathcal{O}_{\mathrm{SM}}\mathcal{O}_\mathcal{U},
\end{equation}
where $d_\mathcal{U}$ and $d_{SM}$ are the scaling dimensions of the unparticle operator and the SM particle operator, respectively.

Now we consider the interactions between unparticles with fermions, specifically, electrons and nucleons.
At low energies, nucleons can be described by a heavy fermion field, and thus the leading interactions in the effective field theory~\cite{Liao2007PRL} can be generalized as
\begin{equation}
\label{interLag}
\mathcal{L}_{\mathrm{int}} =\sum_{i=e,N}(C_S \bar{\psi}_i \psi_i  + C_P \bar{\psi}_i i \gamma_5 \psi_i)\Phi_\mathcal{U}+ (C_V \bar{\psi}_i \gamma_\mu \psi_i +C_A \bar{\psi}_i \gamma_\mu\gamma_5 \psi_i)X_\mathcal{U}^\mu,
\end{equation}
Here $\psi_e$ ($\psi_N$) is the electron (nucleon) field, and $\Phi_\mathcal{U}$ and $X_\mathcal{U}^\mu$ stand for the fields of scalar and vector unparticles, respectively, with  $C_{S,P,V,A}$ being the corresponding coupling constants.
They can be parameterized by $C_{i}=c_{i}\Lambda_{i}^{1-d_\mathcal{U}}$, where $c_i$ and $\Lambda_i$ are unknown dimensionless numbers and energy scales, respectively.
A simple choice is to put $\Lambda_i = 1$ TeV and constrain $c_i$, because they can be easily converted into each other~\cite{Liao2007PRL}.

\begin{figure}[htbp]
	\centering
    \includegraphics[width=0.3\textwidth]{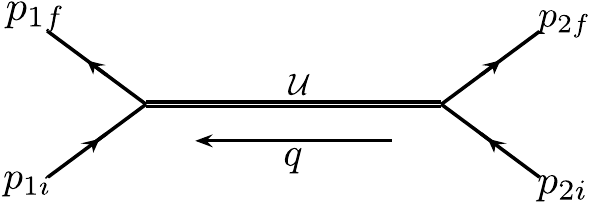}
	\caption{Feynman diagram of scattering between fermions from unparticle exchange. For scalar, pseudoscalar, vector, and pseudovector interaction vertex, 1, $i\gamma_5$, $\gamma_\mu$, and $\gamma_\mu\gamma_5$ should be attached, respectively. $q$ represents the transferred four-momentum. In the nonrelativistic limit, $q$ is replaced by the three-momentum $\vec{q}$.}
	\label{fig:Feydiag}
\end{figure}
The scattering procedure in the nonrelativistic limit can be described by a tree level Feynman diagram, shown in Fig.~\ref{fig:Feydiag}.
From the Lagrangian (\ref{interLag}), six types of interaction, i.e. SS, SP, PP, VV, VA, and AA, can be constructed at tree level.
In the laboratory frame, we consider the elastic scattering with $q_0=0$ and use notations
\begin{equation}
\label{notations}
\begin{split}
\vec q &= \vec p_{1f}- \vec p_{1i}= \vec p_{2i}- \vec p_{2f}, \\
\vec p_1&=\frac{\vec p_{1f} + \vec p_{1i}}{2}, \\
\vec p_2&=\frac{\vec p_{2f} + \vec p_{2i}}{2},
\end{split}
\end{equation}
where $\vec q, \vec p_1, \vec p_2$ are three-momentum components of the corresponding four-momentum.
The scalar unparticle propagator takes the form~\cite{Georgi2007PRL,Cheung2007PRD}
\begin{equation}
D(q^2)=i\frac{A_{d_\mathcal{U}}}{2}\frac{1}{\sin(d_\mathcal{U}\pi)}(-q^2-i\epsilon)^{d_\mathcal{U}-2},
\end{equation}
where $A_{d_\mathcal{U}}=\frac{16\pi^{5/2}}{(2\pi)^{2d_\mathcal{U}}}\frac{\Gamma(d_\mathcal{U}+1/2)}{\Gamma(d_\mathcal{U}-1)\Gamma(2d_\mathcal{U})}$ is the normalization factor.
The vector unparticle propagator $D^{\mu\nu}$ can be obtained by further attaching the spin structure $-g^{\mu\nu}+(1-\xi)q^\mu q^\nu/q^2$.
The six types of amplitude are given by
\begin{equation}
\label{amplitudes}
   \begin{split}
     \mathcal{M}_{SS}&=iC^2_S\bar u(p_{2f})u(p_{2i})D(q^2)\bar u(p_{1f})u(p_{1i}) ,\\
     \mathcal{M}_{SP}&=iC_S(iC_P)\bar u(p_{2f})u(p_{2i})D(q^2)\bar u(p_{1f})\gamma_5u(p_{1i}) ,\\
     \mathcal{M}_{PP}&=i(iC_P)^2\bar u(p_{2f})\gamma_5u(p_{2i})D(q^2)\bar u(p_{1f})\gamma_5u(p_{1i}) ,\\
     \mathcal{M}_{VV}&=iC^2_V\bar u(p_{2f})\gamma_\mu u(p_{2i})D^{\mu\nu}(q^2)\bar u(p_{1f})\gamma_\nu u(p_{1i}) ,\\
     \mathcal{M}_{VA}&=iC_VC_A\bar u(p_{2f})\gamma_\mu u(p_{2i})D^{\mu\nu}(q^2)\bar u(p_{1f})\gamma_\nu \gamma_5u(p_{1i}) ,\\
     \mathcal{M}_{AA}&=iC_A^2\bar u(p_{2f})\gamma_\mu \gamma_5u(p_{2i})D^{\mu\nu}(q^2)\bar u(p_{1f})\gamma_\nu \gamma_5u(p_{1i}),\\
   \end{split}
 \end{equation}
where $u(p)$ is the Dirac spinor.
The amplitude in the nonrelativistic limit reads
\begin{equation}
  \mathcal{A}=\prod_i\frac{1}{\sqrt{2E_i}}\prod_f\frac{1}{\sqrt{2E_f}}\mathcal{M},
\end{equation}
where $i$ ($f$) denotes the index of initial (final) particles, and the coefficients come from the normalization condition.
The potential in coordinate space can be calculated through the Fourier transformation,
\begin{equation}
\label{transform}
V(\vec{r})=-\int\frac{d\vec{q}^3}{(2\pi)^3}\mathcal{A}(\vec{q},\vec{p})e^{i\vec{q}\cdot\vec{r}}.
\end{equation}
The six types of coordinate-space potential keeping terms up to $\mathcal{O}(m^{-3})$ are as follows.
\begin{equation}
\label{SS}
    V_{SS}(r) = -C_S^2\frac{A_{d_\mathcal{U}}}{4\pi^2}r^{1-2d_\mathcal{U}}[\Gamma(2d_\mathcal{U}-2)(1-\frac{\vec p_1^2}{2m_1^2}-\frac{\vec p_2^2}{2m_2^2})
   +\frac{\Gamma(2d_\mathcal{U})}{(2d_\mathcal{U}-2)}(\frac{\vec\sigma_1\cdot(\hat{r}\times\vec p_1)}{4m_1^2r}-\frac{\vec \sigma_2\cdot(\hat{r}\times\vec p_2)}{4m_2^2r})],
\end{equation}
\begin{equation}
\label{SP}
    V_{SP}(r)= -C_SC_P\frac{A_{d_\mathcal{U}}}{4\pi^2}\frac{\Gamma(2d_\mathcal{U})}{2(2d_\mathcal{U}-2)m_1 r^{2d_\mathcal{U}}}\vec\sigma_1\cdot\hat{r},
\end{equation}
\begin{equation}
\label{PP}
    V_{PP}(r) = C_P^2\frac{A_{d_\mathcal{U}}}{4\pi^2}\Gamma(2d_\mathcal{U})\frac{(1+2d_\mathcal{U})(\vec\sigma_1\cdot\hat r)(\vec\sigma_2\cdot\hat r)-\vec\sigma_1\cdot\vec\sigma_2}{4(2d_\mathcal{U}-2)m_1m_2 r^{2d_\mathcal{U}+1}},
\end{equation}
\begin{equation}
\label{VV}
\begin{split}
    V_{VV}(r) = &C_V^2\frac{A_{d_\mathcal{U}}}{4\pi^2}r^{1-2d_\mathcal{U}}[(1-\frac{\vec p_1\cdot\vec p_2}{m_1m_2})\Gamma(2d_\mathcal{U}-2)+\Gamma(2d_\mathcal{U})(\frac{1}{8m_1^2r^2}+\frac{1}{8m_2^2r^2})\\
    &+\frac{\Gamma(2d_\mathcal{U})}{2d_\mathcal{U}-2}\vec\sigma_1\cdot\hat{r}\times(\frac{\vec p_2}{2m_1m_2r}-\frac{\vec p_1}{4m_1^2r}) \\
    &+\frac{\Gamma(2d_\mathcal{U})}{2d_\mathcal{U}-2}\vec\sigma_2\cdot\hat{r}\times(\frac{\vec p_2}{4m_2^2r}-\frac{\vec p_1}{2m_1m_2r})\\
    &-\Gamma(2d_\mathcal{U})\frac{(2d_\mathcal{U}+1)(\vec\sigma_1\cdot\hat r)(\vec\sigma_2\cdot\hat r)+(1-2d_\mathcal{U})(\vec\sigma_1\cdot\vec\sigma_2)}{4(2d_\mathcal{U}-2)m_1m_2r^2}],
    \end{split}
\end{equation}
\begin{equation}
\label{VA}
    V_{VA}(r)=C_VC_A\frac{A_{d_\mathcal{U}}}{4\pi^2}r^{1-2d_\mathcal{U}}[\Gamma(2d_\mathcal{U}-2)(\frac{\vec\sigma_1\cdot\vec p_1}{m_1}-\frac{\vec\sigma_1\cdot\vec p_2}{m_2})-\frac{\Gamma(2d_\mathcal{U})}{2d_\mathcal{U}-2}\frac{(\vec\sigma_2\times\hat{r})\cdot\vec\sigma_1}{2m_2r}],
\end{equation}
\begin{equation}
\label{AA}
  \begin{split}
    V_{AA}(r)= & C_A^2\frac{A_{d_\mathcal{U}}}{4\pi^2}r^{1-2d_\mathcal{U}}\{\Gamma(2d_\mathcal{U}-2)(\frac{\vec p_1^2}{2m_1^2}+\frac{\vec p_2^2}{2m_2^2}-1)\vec\sigma_1\cdot\vec\sigma_2\\
    &+\Gamma(2d_\mathcal{U}-2)[\frac{(\vec\sigma_1\cdot\vec p_1)(\vec\sigma_2\cdot\vec p_2)}{m_1m_2}-\frac{(\vec\sigma_1\cdot\vec p_1)(\vec\sigma_2\cdot\vec p_1)}{2m_1^2}
    -\frac{(\vec\sigma_1\cdot\vec p_2)(\vec\sigma_2\cdot\vec p_2)}{2m_2^2}]\\
    &-\Gamma(2d_\mathcal{U})(\frac{1}{8m_1^2}+\frac{1}{8m_2^2})\frac{(1+2d_\mathcal{U})(\vec\sigma_1\cdot\hat r)(\vec\sigma_2\cdot\hat r)-\vec\sigma_1\cdot\vec\sigma_2}{(2d_\mathcal{U}-2)r^2}\\
    &+\frac{\Gamma(2d_\mathcal{U})}{2d_\mathcal{U}-2}[\frac{\vec\sigma_2\cdot(\vec p_1\times\hat{r})}{4m_1^2r}-\frac{\vec\sigma_1\cdot(\vec p_2\times\hat{r})}{4m_2^2r}]\},
  \end{split}
\end{equation}
where the gauge of $\xi = 1$ is adopted~\cite{Liao2007PRL, Luo2008PLB}.
The main derivation of these potentials, with details for the SS type (\ref{SS}), is given in Appendix~\ref{appendA}.
$\vec\sigma_i$, $m_i$, and $\vec p_i$ are the spin, mass, and momentum of the fermion $i$, respectively.
$\hat{r}$ is a unit vector pointing from particle 1 to particle 2, and $r$ is the distance between them.
Potentials (\ref{SS}), (\ref{PP}), (\ref{VV}), and (\ref{AA}) generated from two same vertexes are invariant under the permutation of particle index.
When dealing with identical particles, potentials (\ref{SP}) and (\ref{VA}) should have terms that exchange particle index 1 and 2 added to them.
This leads to the following expressions:
\begin{equation}
\label{SP2}
    V_{SP}(r) =-C_SC_P\frac{A_{d_\mathcal{U}}}{4\pi^2}\frac{\Gamma(2d_\mathcal{U})}{2(2d_\mathcal{U}-2)mr^{2d_\mathcal{U}}}(\vec\sigma_1-\vec\sigma_2)\cdot\hat{r},
\end{equation}
\begin{equation}
\label{VA2}
    V_{VA}(r)=C_VC_A\frac{A_{d_\mathcal{U}}}{4\pi^2}r^{1-2d_\mathcal{U}}[\Gamma(2d_\mathcal{U}-2)\frac{1}{m}(\vec\sigma_1-\vec\sigma_2)\cdot(\vec p_1-\vec p_2)
     -\frac{\Gamma(2d_\mathcal{U})}{2d_\mathcal{U}-2}\frac{(\vec\sigma_2\times\hat{r})	\cdot\vec\sigma_1}{m r}].
\end{equation}
These expressions are such that they do not vanish when the wave function is antisymmetric.

The potentials we derived contain terms up to $\mathcal{O}(m^{-3})$ because higher-order effects are largely suppressed by the mass and are therefore irrelevant for observations in low-energy experiments.
$\vec{r}$ and $\vec{p}$ should be treated as vectors/operators for macroscopic/microscopic systems, e.g., the term $\vec{\sigma}_1\cdot(\hat{r}\times\vec{p}_1)$ should be replaced by $(\vec{\sigma}_1\times\hat{r})\cdot\vec{p}_1+\vec{p}_1\cdot(\vec{\sigma}_1\times\hat{r})$ for the interaction of SS type at the atomic scale~\cite{Fadeev2019PRA}.
One can put $\vec{p}_1=-\vec{p}_2$ to obtain the potentials in the center-of-mass frame.
It is important to note that these potentials exhibit a strong dependence on the continuous scaling dimension $d_\mathcal{U}$.
This is due to the exponential factor of $r$, which dominates the behavior of the potentials at small distances $r$.
The significant dependence on $d_\mathcal{U}$ has been highlighted by studying the unparticle contribution to the ground-state energy of the hydrogen atom~\cite{Wondrak2016PLB}.
In the following discussion, for convenience, we take a range of $d_\mathcal{U} \in (1,2)$~\cite{Georgi2007PLB,Liao2007PRL}.
This range could be suitably modified or adjusted in certain cases, as discussed in Ref.~\cite{Grinstein2008PLB}.

\section{Constraints from existing experiments}

Potentials mediated by unparticles are particularly special as they follow a non-Yukawa $1/r^\alpha$  relationship, where $\alpha$ is a positive real number greater than 1 and not necessarily an integer, which is a unique characteristic of these potentials.
Some other types of non-Yukawa spin-dependent interactions were recently studied in Ref.~\cite{Costantino2020JHEP}.
This could be viewed as a possible origin of the deviation from the gravitational inverse square law (ISL).
Previously, such kind of deviation was modeled by including a Yukawa type or $1/r^N (N=2,3,\cdots)$ type potential, mediated by a finite mass particle or $N$ massless particles, into the Newtonian potential~\cite{Adelberger2007PRL}.
If the violation of ISL is considered due to unparticle effects, constraints on unparticle could be obtained.
For instance, treating the nucleon as baryon number current, Ref.~\cite{Deshpande2008PLB} estimated the limits on the long range interactions induced by vector unparticle exchange.
Similarly, the bounds on such interactions have been obtained from geodetic and frame-dragging measurements by considering the vector unparticle coupling to the electronic (leptonic) and the nucleonic (baryonic) currents~\cite{Poddar2022EPJC}.
Here, describing the nucleon as a fermion field, we use the data in Ref.~\cite{Adelberger2007PRL} to constrain $c_V$, with $\Lambda_\mathcal{U} = 1$~TeV, $d_\mathcal{U} = 1.5$ and $2.0$.
The details to derive the constraint are given in Appendix~\ref{appendB}.
The results are tabulated in Table~\ref{table1}, alongside the constraint in Ref.~\cite{Chang2011EPJC}.
The obtained new constraint for $d_\mathcal{U} = 1.5$ improves the previous limit by as much as 9 orders of magnitude.
\begin{table}[htbp]
\centering
\caption{Constraints on the coupling constant $c_V$ of VV type interaction between nucleons (\ref{VV}) at the leading term for $d_\mathcal{U} = 1.5$ and $2.0$ at the $68\%$ confidence level. The energy scale $\Lambda_{\mathcal{U}}$ is fixed as $1$~TeV.}
\begin{tabular}{ccc}
	\hline
    $d_\mathcal{U}$ & 1.5 & 2.0  \\
	\hline
    $c_V$ & $1.7\times 10^{-10}$ & $1.8\times 10^{-2}$  \\
    $c_V$~\cite{Chang2011EPJC} & $3.7\times 10^{-1}$ & $-$ \\
	\hline
\label{table1}
\end{tabular}
\end{table}

It is worth noting that the potentials of SP and VA types, originating from couplings of different types of vertex, violate $P$, $T$ and $C$, $P$ symmetries, respectively.
The amount of $CP$ violation permitted by the SM has been recognized to be inadequate for explaining the observed asymmetry between matter and antimatter in the Universe.
Therefore, the search for novel sources of $CP$ and $T$ violations has become a crucial frontier in modern physics research~\cite{Dine2003RMP}.
A particularly relevant example of the $P$, $T$ violation effect is the electric dipole moments (EDMs), which could be induced by $V_{SP}$ in atoms due to the mixture of opposite-parity eigenstates.
Based on the precise measurement of the EDM of $^{199}$Hg atom~\cite{Graner2016PRL,Graner2017Erratum}, the constraint on coupling constants of $V_{SP}$ for $d_\mathcal{U} = 1.0$ is evaluated as
 \begin{equation}
 	|c_S^ec_P^N|=7.0\times 10^{-17},
 \end{equation}
at the $95\%$ confidence level.
The details to obtain this constraint can be also found in Appendix~\ref{appendB}.
Previously, the effects of $CP$ violation from unparticle physics were only explored in collider physics~\cite{Chen2007PRD,Ettefaghi2017PRD}.
Here, the exchange of unparticles through $V_{SP}$ and $V_{VA}$ can offer an additional approach to probe the effects of discrete symmetry violation in precision measurement experiments.

\section{Conclusion and Discussion }
\label{discuss}

In summary, six types of coordinate-space potential between fermions from unparticle exchange are derived under the nonrelativistic limit, which could be probed in precision measurement experiments.
Previous studies have focused on parity-even interactions, such as SS, PP, VV, and AA.
We propose new interactions of the SP and VA type, which will not conserve parity.
It is important to note that parity conservation is not always a fundamental principle in modern physics.
Properties of these potentials are discussed.
Based on existing experimental data, constraints on some coupling constants are estimated.
By examining the unparticle effects in the violation of the gravitational inverse square law, our limit on the vector unparticle coupling with fermions, $c_V = 1.7\times 10^{-10}$, improves the previous constraint by as much as 9 orders of magnitude.
By using the upper limit for the electric dipole moment of the $^{199}$Hg atom, we also constrain the coupling constants of SP type interaction from unparticle exchange.

Finally, we point out that more constraints could be obtained considering the spin dependence of most potentials.
The spin-dependent terms can be written as $\vec\sigma\cdot\vec{B}_{\mathrm{pseudo}}$, where $\vec{B}_{\mathrm{pseudo}}$ can be viewed as a kind of pseudo-magnetic field.
Such pseudo-magnetic field can be sensed by spin-polarized quantum sensor, e.g. the atom magnetometer.
The state-of-the-art magnetometer working in Spin Exchange Relaxation Free region has reached the sensitivity of $\sim\mathrm{fT}/\sqrt{\mathrm{Hz}}$, i.e. the energy resolution of $\sim 10^{-23}~\mathrm{eV}/\sqrt{\mathrm{Hz}}$~\cite{Budker2007NatPhys}.
This kind of techniques has been widely used in the table-top new physics research, and many progresses have been recently reported~\cite{Kim2018PRL,Terrano2021QST,Su2021SciAdv,Wu2022PRL,Wei2022NC,Xiao2023PRL}.
These findings can aid in constraining spin-dependent interactions that are mediated by unparticles, and ongoing work is being conducted in this area.

\begin{acknowledgments}
We acknowledge supports from the National Natural Science Foundation of China under grants U2230207 and U2030209 and the National Key Program for Research and Development of China under grants 2020YFA0406001 and 2020YFA0406002. We thank Dr. Liao for the helpful discussions and for providing valuable notes on the relevant subjects.
\end{acknowledgments}

\appendix
\section{Derivation of potentials in coordinate space}\label{appendA}

First some useful identities are listed as follows.
 \begin{equation}
 \label{id2}
 \begin{split}
   & (\vec\sigma\cdot\vec A)(\vec\sigma\cdot\vec B) = \vec A\cdot\vec B+i\vec\sigma\cdot(\vec A\times\vec B) \\
   & (\vec\sigma_1\times\vec A)\cdot(\vec\sigma_2\times\vec B) =(\vec A\cdot\vec B)(\vec\sigma_1\cdot\vec\sigma_2)-(\vec\sigma_2\cdot\vec A)(\vec\sigma_1\cdot\vec B)\\
   & (\vec\sigma\cdot\vec A)\sigma_i(\vec\sigma\cdot\vec B)=(\vec \sigma\cdot\vec A)\vec B_i+i(\vec B\times\vec A)_i-(\vec A\cdot\vec B)\vec\sigma_i+(\vec\sigma\cdot\vec B)\vec A_i\\
  \end{split}
 \end{equation}
\begin{equation}
\label{d1}
	\partial_i\partial_j r^{1-2d}=(1-2d)(-1-2d)r^{-3-2d}x_ix_j+(1-2d)r^{-1-2d}\delta_{ij}
\end{equation}
\begin{equation}
\label{d2}
	\nabla r^{1-2d}=(1-2d)r^{-2d}\hat{r}
\end{equation}
\begin{equation}
 \label{f1}
	\int\frac{d\vec q^3}{(2\pi)^3}\frac{1}{(\vec q^2)^k}e^{i\vec q\cdot \vec r}=\frac{1}{2\pi^2}r^{2k-3}\sin(k\pi)\Gamma(2-2k)
\end{equation}
\begin{equation}
\label{f2}
	 \int\frac{d\vec q^3}{(2\pi)^3}\frac{\vec q}{(\vec q^2)^k}e^{i\vec q\cdot \vec r}=(-i\nabla)\frac{1}{2\pi^2}r^{2k-3}\sin(k\pi)\Gamma(2-2k)
\end{equation}
In the following, the natural units $\hbar =c=1$, metric signature $(+\ ,-\ ,-\ ,-)$, and $\gamma_\mu$ matrices in the Dirac representation are adopted.
The Dirac spinor in Eq.~(\ref{amplitudes}) reads
 \begin{equation}
  u(p)=N\left(
  \begin{array}{c}
  \chi_s \\
  \frac{\vec\sigma\cdot \vec p}{E+m}\chi_s
\end{array}
\right),
\end{equation}
where $\chi_s$ is the spin wave function with $s=0$ or $1$, $N=\sqrt{E+m}$ is the normalization factor, and $E$, $\vec p$, and $m$ is the energy, three-momentum, and rest mass of the particle, respectively.
Taking the scalar type spinor product as an example, we have
\begin{equation}
 \label{S}
 \begin{split}
   \bar{u}(p_{1f})u(p_{1i})&=u(p_{1f})^\dagger\left(
   \begin{array}{cc}
   	1 & 0 \\
   	0 & -1
  \end{array}
   \right)u(p_{1i})\\
   &=N_1^2(1-\frac{(\vec\sigma_1\cdot\vec p_{1f})(\vec\sigma_1\cdot\vec p_{1i})}{(E_1+m_1)^2})\\
   &=N_1^2(1-\frac{\vec p_{1f}\cdot\vec p_{1i}+i\vec\sigma_1\cdot(\vec p_{1f}\times\vec p_{1i})}{(E_1+m_1)^2})\\
   &=N_1^2(1-\frac{\vec p_{1i}^2+\vec p_{1i}\cdot\vec q+i\vec\sigma_1\cdot(\vec q\times\vec p_{1i})}{(E_1+m_1)^2})\\
   &=N_1^2(1-\frac{\vec p_1^2-\vec q^2/4+i\vec\sigma_1\cdot(\vec q\times\vec p_1)}{(E_1+m_1)^2}),\\
  \end{split}
\end{equation}
where $\vec p_{1i}$ ($\vec p_{1f}$) refers to the three-momentum of the initial (final) state of particle 1.
In an elastic scattering process, $E_{1i}=E_{1f}=E_1$.
As defined in Eq.~(\ref{notations}), $\vec p_1=(\vec p_{1i}+\vec p_{1f})/2$ and $\vec q=\vec p_{1f}-\vec p_{1i}$ are the average and transferred three-momenta, respectively.
Other spinor products, listed below, can be calculated in a similar way.
 \begin{equation}
   \bar{u}(p_{1f})\gamma_5u(p_{1i}) =N_1^2\frac{\vec\sigma_1\cdot(\vec p_{1i}-\vec p_{1f})}{E_1+m_1}
   =-N_1^2\frac{\vec\sigma_1\cdot\vec q}{E_1+m_1}
\end{equation}
\begin{equation}
   \bar{u}(p_{1f})\gamma_0u(p_{1i})=N_1^2(1+\frac{\vec p_1^2-\vec q^2/4+i\vec\sigma_1\cdot(\vec q\times\vec p_1)}{(E_1+m_1)^2})
\end{equation}
\begin{equation}
  \bar{u}(p_{1f})\gamma_iu(p_{1i})=N_1^2(\frac{2(\vec p_1)_i}{E_1+m_1}+\frac{i(\vec\sigma_1\times\vec q)_i}{E_1+m_1})
\end{equation}
\begin{equation}
   \bar{u}(p_{1f})\gamma_0\gamma_5u(p_{1i})=N_1^2\frac{2\vec\sigma_1\cdot\vec p_1}{E_1+m_1}
\end{equation}
\begin{equation}
   \bar{u}(p_{1f})\gamma_i\gamma_5u(p_{1i})=
   N_1^2((\vec\sigma_1)_i+\frac{2(\vec\sigma_1\cdot\vec p_1)(\vec p_1)_i-(\vec\sigma_1\cdot\vec q)(\vec q_1)_i/2+i(\vec p_1\times\vec q)_i-(\vec p_1^2-\vec q^2/4)(\vec\sigma_1)_i}{(E_1+m_1)^2})
\end{equation}

The nonrelativistic scattering amplitude for the interaction of SS type reads
\begin{equation}
\begin{split}
	\mathcal{A}_{SS}&=\frac{1}{4E_1E_2}\mathcal{M}_{SS}\\
	&=\frac{N_1^2N_2^2}{4E_1E_2}iC^2_S\frac{A_{d_\mathcal{U}}}{2\sin(\pi d_\mathcal{U})}\frac{i}{(-q^2-i\epsilon)^{2-d_\mathcal{U}}}
	(1-\frac{\vec p_1^2-\vec q^2/4+i\vec\sigma_1\cdot(\vec q\times\vec p_1)}{(E_1+m_1)^2})(1-\frac{\vec p_2^2-\vec q^2/4-i\vec\sigma_2\cdot(\vec q\times\vec p_2)}{(E_2+m_2)^2}).
\end{split}
\end{equation}
By expanding the energy as $E=\sqrt{m^2+\vec p^2}=m+\frac{\vec p^2}{2m}+\mathcal{O}(m^{-2})$, one has
\begin{equation}
\label{Ass}
    \mathcal{A}_{SS} =-\frac{A_{d_\mathcal{U}}}{2\sin(\pi d_\mathcal{U})}C_S^2\frac{1}{(\vec q^2)^{2-d_\mathcal{U}}}(1-\frac{2\vec p_1^2+i\vec\sigma_1\cdot(\vec q\times\vec p_1)}{4m_1^2}
    -\frac{2\vec p_2^2-i\vec\sigma_2\cdot(\vec q\times\vec p_2)}{4m_2^2}).
\end{equation}
The nonrelativistic amplitudes for other interactions are listed as follows.
\begin{equation}
    \mathcal{A}_{SP} = i\frac{A_{d_\mathcal{U}}}{2\sin(\pi d_\mathcal{U})}C_SC_P\frac{1}{(\vec q^2)^{2-d_\mathcal{U}}}\frac{\vec\sigma_1\cdot\vec q}{2m_1},
\end{equation}
\begin{equation}
    \mathcal{A}_{PP} =-\frac{A_{d_\mathcal{U}}}{2\sin(\pi d_\mathcal{U})}C_P^2\frac{1}{(\vec q^2)^{2-d_\mathcal{U}}}\frac{(\vec\sigma_1\cdot\vec q)(\vec\sigma_2\cdot\vec q)}{4m_1m_2},
\end{equation}
\begin{equation}
\begin{split}
    \mathcal{A}_{VV}= & \frac{A_{d_\mathcal{U}}}{2\sin(\pi d_\mathcal{U})}C_V^2\frac{1}{(\vec q^2)^{2-d_\mathcal{U}}}(1+\frac{-\vec q^2/2+i\vec\sigma_1\cdot(\vec q\times\vec p_1)}{4m_1^2}+\frac{-\vec q^2/2-i\vec\sigma_2\cdot(\vec q\times\vec p_2)}{4m_2^2}\\
    &-\frac{4\vec p_1\cdot\vec p_2+(\vec\sigma_1\times\vec q)(\vec\sigma_2\times\vec q)+i(2\vec p_2\cdot(\vec\sigma_1\times\vec q)-2\vec p_1\cdot(\vec\sigma_2\times\vec q))}{4m_1m_2}),
\end{split}
\end{equation}
\begin{equation}
    \mathcal{A}_{VA} =\frac{A_{d_\mathcal{U}}}{2\sin(\pi d_\mathcal{U})}C_VC_A\frac{1}{(\vec q^2)^{2-d_\mathcal{U}}}(\frac{\vec\sigma_1\cdot\vec p_1}{m_1}-\frac{\vec\sigma_1\cdot\vec p_2}{m_2}+\frac{i(\vec\sigma_2\times\vec q)\cdot\vec\sigma_1}{2m_2}),
\end{equation}
\begin{equation}
\begin{split}
    \mathcal{A}_{AA}=& \frac{A_{d_\mathcal{U}}}{2\sin(\pi d_\mathcal{U})}C_V^2\frac{1}{(\vec q^2)^{2-d_\mathcal{U}}}\{-\vec\sigma_1\cdot\vec\sigma_2+\frac{(\vec\sigma_1\cdot\vec p_1)(\vec\sigma_2\cdot\vec p_2)}{m_1m_2}\\
    &-\frac{2(\vec\sigma_1\cdot\vec p_1)(\vec\sigma_2\cdot\vec p_1)-(\vec\sigma_1\cdot\vec q)(\vec\sigma_2\cdot\vec q)/2+i\vec\sigma_2\cdot(\vec p_1\times\vec q)-2\vec p_1^2\vec\sigma_1\cdot\vec\sigma_2}{4m_1^2}\\
    &-\frac{2(\vec\sigma_1\cdot\vec p_2)(\vec\sigma_2\cdot\vec p_2)-(\vec\sigma_2\cdot\vec q)(\vec\sigma_1\cdot\vec q)/2-i\vec\sigma_1\cdot(\vec p_2\times\vec q)-2\vec p_2^2\vec\sigma_1\cdot\vec\sigma_2}{4m_2^2}\}.\\
  \end{split}
\end{equation}
Applying the Fourier transformation (\ref{transform}) to the amplitude (\ref{Ass}), one finally reaches
\begin{equation}
\begin{aligned}
  V_{SS}(\vec{r})= & -\int\frac{d^3\vec q}{(2\pi)^3}\mathcal{A}_{SS}(\vec{q},\vec{p})e^{i\vec q\cdot \vec r}  \\
  = & \frac{A_{d_\mathcal{U}}}{2\sin(\pi d_\mathcal{U}){(2\pi)^3}}C_S^2 \{\int\frac{1}{(\vec q^2)^{2-d_\mathcal{U}}}(1-\frac{\vec p_1^2}{2m_1^2}-\frac{\vec p_2^2}{2m_2^2})e^{i\vec q\cdot \vec r} d^3\vec q \\
  &-\frac{i\vec\sigma_1\cdot(\vec q\times\vec p_1)}{4m_1^2}e^{i\vec q\cdot \vec r}d^3\vec q+\frac{i\vec\sigma_2\cdot(\vec q\times\vec p_2)}{4m_2^2}e^{i\vec q\cdot \vec r}d^3\vec q\}\\
  =& -\frac{A_{d_\mathcal{U}}}{4\pi^2}C_S^2 \Gamma(2d_\mathcal{U}-2) r^{1-2d_\mathcal{U}}\{(1-\frac{\vec p_1^2}{2m_1^2}-\frac{\vec p_2^2}{2m_2^2})-\frac{\vec\sigma_1\cdot(\nabla\times\vec p_1)}{4m_1^2}+\frac{\vec\sigma_2\cdot(\nabla\times\vec p_2)}{4m_2^2}\}\\
  = & -\frac{A_{d_\mathcal{U}}}{4\pi^2}C_S^2 \Gamma(2d_\mathcal{U}-2) r^{1-2d_\mathcal{U}}\{(1-\frac{\vec p_1^2}{2m_1^2}-\frac{\vec p_2^2}{2m_2^2})-(1-2d_\mathcal{U})(\frac{\vec\sigma_1\cdot(\hat{r}\times\vec p_1)}{4m_1^2r}-\frac{\vec\sigma_2\cdot(\hat{r}\times\vec p_2)}{4m_2^2r})\}\\
  = &  -\frac{A_{d_\mathcal{U}}}{4\pi^2}C_S^2 r^{1-2d_\mathcal{U}} \{\Gamma(2d_\mathcal{U}-2)(1-\frac{\vec p_1^2}{2m_1^2}-\frac{\vec p_2^2}{2m_2^2}) +\frac{\Gamma(2d_\mathcal{U})}{(2d_\mathcal{U}-2)}(\frac{\vec\sigma_1\cdot(\hat{r}\times\vec p_1)}{4m_1^2r}-\frac{\vec\sigma_2\cdot(\hat{r}\times\vec p_2)}{4m_2^2r})\},
\end{aligned}
\end{equation}
with the use of Eqs.~(\ref{d2})-(\ref{f2}).
This is exactly the potential (\ref{SS}) with an explicit inclusion of the common factor and the coupling strength.
Following a similar procedure, one can obtain the coordinate-space potentials (\ref{SP})-(\ref{AA}).

\section{Constraint on VV and SP type interactions}
\label{appendB}
In the torsion-balance experiment, the power-law potential,
\begin{equation}
	V^k_{ab}(r)=G\frac{M_1M_2}{r}\beta_k(\frac{1\ \mathrm{mm}}{r})^{k-1},
\end{equation}
was considered as a form of possible breakdown of the ISL, and some constraints on it was obtained \cite{Adelberger2007PRL}.
The total power-law interaction between the molybdenum pendulum and the tantalum attractors is
\begin{equation}
    \label{pl}
    U^k_{ab}(r)=\int \int \rho_1\rho_2 V^k_{ab}(r)dV_1dV_2,
\end{equation}
where $\rho_1=0.01028\ \mathrm{g/mm^3}$ and $\rho_2=0.01669\ \mathrm{g/mm^3}$ are the corresponding mass densities.
If VV type interaction is also thought of possible form, the corresponding interaction reads
\begin{equation}
    \label{uvv}
    U_{VV_0}(r)=\int \int \rho_{N1}\rho_{N2} V_{VV_0}(r)dV_1dV_2,
\end{equation}
where $\rho_{N_1}=6.45\times 10^{19}\ \mathrm{mm^{-3}}$ and $\rho_{N_2}=5.55\times 10^{19}\ \mathrm{\mathrm{mm^{-3}}}$ are the nucleon number densities of molybdenum and tantalum, respectively.
Here, $V_{VV_0}(r)$ is the first term of Eq.~(\ref{VV}), which can be expressed as
\begin{equation}
\label{VVSI}
    V_{VV_0}(r)=c_V^2\Lambda^{2-2d_\mathcal{U}}\frac{16\pi^{5/2}}{(2\pi)^{2d_\mathcal{U}}4\pi^2}\frac{\Gamma(d_\mathcal{U}+1/2)\Gamma(2d_\mathcal{U}-2)}{\Gamma(d_\mathcal{U}-1)\Gamma(2d_\mathcal{U})}(\hbar c)^{2d_\mathcal{U}-1}(1\ \mathrm{mm})^{1-2d_\mathcal{U}}(\frac{1\ \mathrm{mm}}{r})^{2d_\mathcal{U}-1},
\end{equation}
in the international system (SI) of units.
By comparing Eqs.~(\ref{pl}) and (\ref{uvv}), one obtains the constraint
\begin{equation}
    c_V=\sqrt{G\rho_1\rho_2\beta_k/(\Lambda^{2-2d_\mathcal{U}}\frac{16\pi^{5/2}}{(2\pi)^{2d_\mathcal{U}}4\pi^2}\frac{\Gamma(d_\mathcal{U}+1/2)\Gamma(2d_\mathcal{U}-2)}{\Gamma(d_\mathcal{U}-1)\Gamma(2d_\mathcal{U})}(\hbar c)^{2d_\mathcal{U}-1}(1\ \mathrm{mm})^{2-2d_\mathcal{U}}\rho_{N_1}\rho_{N_2})},
\end{equation}
when $k=2d_\mathcal{U}-1$.

The most stringent constraint on the EDM of $^{199}$Hg atom \cite{Graner2016PRL,Graner2017Erratum} is applyed to constrain the axion mediated $P, T$ violating interaction~\cite{Dzuba2018PRD},
\begin{equation}
    \label{ap}
    V(r)=-\frac{g_e^sg_N^p}{8\pi m_N}\boldsymbol{\sigma_N}\cdot \boldsymbol{\nabla}(\frac{e^{-m_ar}}{r})\gamma^0.
\end{equation}
By considering the limit of a small axion mass, the interaction (\ref{ap}) reduces to
\begin{equation}
    \label{lap}
    \lim_{m_a\rightarrow 0}V(r)=\frac{g_e^sg_N^p}{8\pi m_N}\boldsymbol{\sigma_N}\cdot \boldsymbol{\hat{r}}\frac{1}{r^2}\gamma^0,
\end{equation}
where $m_N$ and $\boldsymbol{\sigma_N}$ are respectively the mass and spin vector of the nucleon, and the Dirac matrix $\gamma^0$ corresponds to the electron.
The SP potential (\ref{SP}) at $d_\mathcal{U}\rightarrow 1$ reads
\begin{equation}
    \label{lup}
    \lim_{d_\mathcal{U}\rightarrow 1}V_{SP}(r)=-\frac{c^e_Sc^N_P}{8\pi m_N}\boldsymbol{\sigma_N}\cdot \boldsymbol{\hat{r}}\frac{1}{r^2}\gamma^0,
\end{equation}
which takes a similar form of the potential (\ref{lap}), and has an appended $\gamma^0$ for the relativistic electron.
By comparing Eqs. (\ref{lap}) and (\ref{lup}), one obtains the constraint on $|c^e_Sc^N_P|$.

%

\end{document}